\documentclass[10pt]{iopart}

\usepackage{siunitx}
\usepackage{amsfonts}
\usepackage{amssymb}
\usepackage{graphicx}
\usepackage[autostyle]{csquotes} 
\usepackage{bm}
\usepackage{subcaption} 
\usepackage{booktabs}
\usepackage{makecell}

\usepackage[switch]{lineno} 
\usepackage[
    colorlinks=true,
    linkcolor=blue,
    citecolor=blue,
    urlcolor=blue,
    pdfauthor={Sascha Rienaecker},
    bookmarksopen=true,
    ]{hyperref}
    
\usepackage[backend=biber,
    bibstyle=numeric-comp,
    sorting=none, 
    citestyle=numeric-comp, %
    maxcitenames=1,
    ]{biblatex}

\AtEveryBibitem{%
\clearfield{url}%
\clearfield{urlyear}%
\clearfield{title}%
}

\renewcommand{\vec}[1]{\bm{#1}}
\newcommand{\ttiny}[1]{\text{\tiny{#1}}} 
\newcommand{\low}[1]{_\mathrm{#1}}

\newcommand{\refeq}[1]{Eq.\,(\ref{eq:#1})}

\newcommand{\refig}[1]{Fig.\,\ref{fig:#1}}
\newcommand{\refsec}[1]{Sec.\,\ref{sec:#1}}

\newcommand{\refappend}[1]{\ref{append:#1}}
\def\documentspath{/home/rienacker/Documents} 

\def\EPSfigpath{\documentspath/EPS24-Salamanca/figures}

\newcommand{\ErxB}{E_r \times B}
\newcommand{\BxgradB}{B \! \times \! \nabla B}

\graphicspath{
              {\EPSfigpath/}
              }

\usepackage[final, authormarkup=none=color, commandnameprefix=ifneeded]{changes} 
\definechangesauthor[name={ReviewerA}, color=red]{A}
\definechangesauthor[name={ReviewerB}, color=violet]{B}

\makeatletter
\newcommand{\mainmatter}{%
\setcounter{footnote}{0}%
\patchcmd{\@makefntext}{\fnsymbol}{\alph}{}{}%
\patchcmd{\@thefnmark}{\fnsymbol}{\alph}{}{}%
\def\@makefnmark{\textsuperscript{\alph{footnote}}}%
}
\makeatother

\begin{document}

\title{Survey of the Edge Radial Electric Field in L-mode TCV Plasmas using Doppler Backscattering}


\author{S. Rienäcker$^1$, P. Hennequin$^1$, L. Vermare$^1$, C. Honoré$^1$, 
        S. Coda$^2$, B. Labit$^2$, B. Vincent$^2$, Y. Wang$^2$, L. Frassinetti$^3$, 
        O. Panico$^{1,4}$, the TCV team\footnote{See author list of B.P. Duval et al 2024 Nucl. Fusion 64 112023} and the EUROfusion Tokamak Exploitation team\footnote{See the author list of E. Joffrin et al 2024 Nucl. Fusion 64 112019}}

\address{$^1$ Laboratoire de Physique des Plasmas (LPP), CNRS, Sorbonne Université, École polytechnique, Institut Polytechnique de Paris, Palaiseau, France}
\address{$^2$ Ecole Polytechnique Fédérale de Lausanne, Swiss Plasma Center, Lausanne, Switzerland}
\address{$^3$ Division of Fusion Plasma Physics, KTH Royal Institute of Technology, Stockholm, Sweden}
\address{$^4$ CEA, IRFM, F-13108 Saint-Paul-lez-Durance, France}

\ead{sascha.rienacker@lpp.polytechnique.fr}
\vspace{10pt}
\begin{indented}
\item[]April 2025
\end{indented}



\begin{abstract}

    A Doppler backscattering (DBS) diagnostic has recently been installed on the \textit{Tokamak à Configuration Variable} (TCV) to facilitate the study of edge turbulence and flow shear in a versatile experimental environment.
    The dual channel V-band DBS system is coupled to TCV's quasi-optical diagnostic launcher, providing access to the upper low-field side region of the plasma cross-section.
    Verifications of the  DBS measurements are presented.
    The DBS equilibrium $v_\perp$ profiles are found to compare favorably with gas puff imaging (GPI) measurements and to the $E_r$ inferred from the radial force balance of the carbon impurity.
    The radial structure of the edge $E_r \times B$ equilibrium flow and its dependencies are investigated across a representative set of L-mode TCV discharges, by varying density, auxiliary heating and magnetic configuration.
\end{abstract}

%
\submitto{\PPCF}
%
%
\ioptwocol

\mainmatter

\section{Introduction}

In the boundary region of tokamak plasmas, a narrow layer with strong $E_r \times B$ velocity shear has broadly been observed experimentally. It is associated with a sharp minimum of the radial electric field, referred to as the $E_r$ \enquote{well}.
Sheared flows are widely recognized as important for the regulation of turbulence and the formation of transport barriers\,\cite{Biglari_1990, Burrell_1997}, allowing for the bifurcation towards regimes of enhanced confinement, such as the low to high confinement mode (L-H) transition.
Meanwhile, the $E_r$ well exhibits a rich phenomenology\,\cite{Meyer_2008, Vermare_2021, Silva_2021, Plank_2023}, and the dominant mechanisms setting its detailed radial structure (and hence the velocity shear) remain poorly understood. 
The precise characterization of $E_r$ serves the validation of transport models and first-principle simulations.
Moreover, it could offer insights into the primary drives underlying the $E_r \! \times \! B$ flow and clarify the interplay between velocity shear and turbulence, in particular at the onset of confinement transitions.

Motivated by these considerations, this paper presents a phenomenological overview of the $E_r$ profile in the \textit{Tokamak à Configuration Variable} (TCV). Taking advantage of the high spatial and temporal resolution offered by microwave diagnostics, specifically Doppler backscattering (DBS), along with TCV's versatility, the edge $E_r$ structure is documented over a diverse range of discharge conditions.

The TCV tokamak\,\cite{Reimerdes_2022}
is a medium-size, carbon-wall tokamak with major radius $R = \SI{0.88}{m}$, 
minor radius $a = \SI{0.25}{m}$, on-axis magnetic field $|B_0| < \SI{1.54}{T}$, plasma current $|I_p| < \SI{1}{MA}$
and discharge duration $\approx \SI{2}{s}$. 
TCV stands out for its uniquely flexible magnetic control capabilities, with a wide range of accessible plasma shapes and vertical positions. 
Furthermore, the signs of $B_0$ and $I_p$ can be chosen freely and independently of each other.
Together with the high power density heating system composed of electron cyclotron resonance heating (ECRH) and neutral beam injection (NBI) with two counter-injecting beams\,\cite{Karpushov_2023}, TCV offers a good opportunity for systematic investigations of $E_r$ and its parameter dependencies.

The radial profiles of the $E_r \! \times \! B$ rotation are measured using a recently installed dual-channel V-band ($50$-$75$\,GHz) DBS system, on loan from the LPP group. DBS studies at TCV have been conducted in the past, using \replaced[id=A]{a similar predecessor\,\cite{Meijere_2014}, or with SPC's in-house DBS system\,\cite{Cabrera_2018}.}{ other systems\,\cite{Meijere_2014, Cabrera_2018}, and}
$E_r$ profiles have also been reconstructed from CXRS measurements\,\cite{Duval_2007, Marini_2017_phdthesis}, but an extended characterization of the edge $E_r$ has not yet been undertaken for this tokamak.

The remainder of this paper is structured as follows: \refsec{DBS_at_TCV} details the DBS system, along with the associated methodology and testing.
The results of the $E_r$ characterization are presented in \refsec{results}.
A summary and conclusions are given in \refsec{conclusion}, while the coordinates and sign conventions are outlined in \refappend{conventions}.

\section{Doppler backscattering at TCV}\label{sec:DBS_at_TCV}


\subsection{DBS working principles}\label{sec:DBS_principles}

\begin{figure}[htbp]
	\centering
		\includegraphics[width=1.0\linewidth]{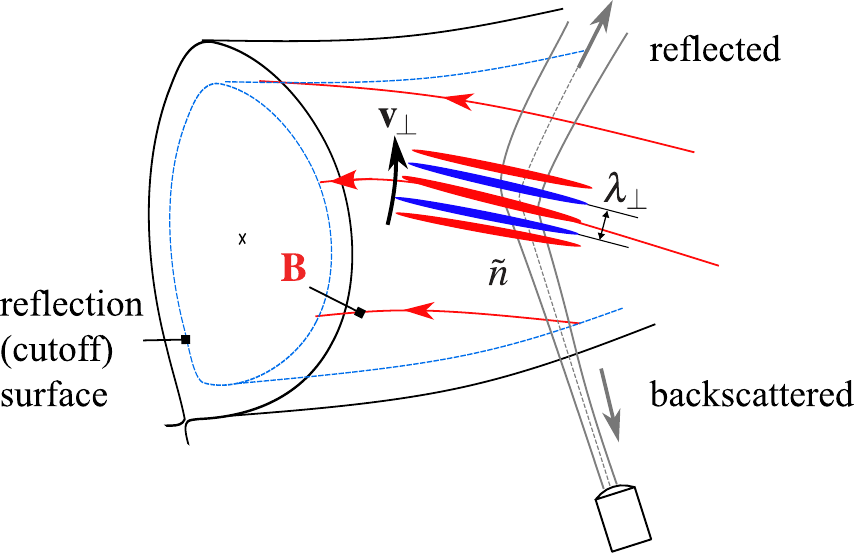}
	  \caption{Sketch of the DBS method: Only the wave backscattered off density fluctuations ($\tilde{n}$), whose size $\lambda_\perp = 2 \pi / k_\perp$ matches the scattering condition, is detected. The scattering efficiency is maximized using near-perpendicular incidence to $\vec{B}$.}\label{fig:DBS_principle}
\end{figure}

Doppler backscattering (DBS)\,\cite{Hirsch_2001, Conway_2004, Hennequin_2006}, also known as Doppler \textit{reflectometry}, is a non-intrusive, lightweight diagnostic that is based on the scattering of an incident microwave beam to infer information on the plasma turbulence and flow. With a suitably chosen frequency and polarization, the incident probing beam encounters a reflecting cutoff layer in the plasma.
By deliberately tilting the beam, see \refig{DBS_principle}, the reflected wave is spatially separated from the one backscattered by density perturbations ($\tilde{n}$). The backscattered wave then travels back to the antenna, where it is detected.

The combination of wave electric field swelling near the cutoff, turbulence anisotropy ($k_\parallel \ll k_\perp$) and the condition for efficient backscattering, known as the Bragg selection rule, ensures that the scattered signal predominantly stems from fluctuations close to the beam turning point and of a given perpendicular spatial scale that satisfies $\vec{k}_\perp = - 2 \vec{k}_i$. Here, $\vec{k}_\perp$ and $\vec{k}_i$ are the wavevectors of the fluctuations and of the incident probing beam (at the turning point), respectively. The turning point location and $\vec{k}_i$ are commonly computed via raytracing methods. The scattering efficiency is maximized for near-perpendicular beam incidence to the local magnetic field, i.e., $\vec{k}_i \perp \vec{B}$. Therefore, the beam is \replaced[id=A]{also tilted}{tilted also} toroidally to roughly match the pitch angle in the probing region.

The selected scattering structures are carried by the local $\ErxB$ plasma flow, thus introducing a Doppler shift $\omega\low{Dop} = k_\perp v_\perp$ in the scattered signal, where $v_\perp$ is the perpendicular velocity of the scatterers in the lab frame.
If the turbulence \textit{intrinsic} velocity (relative to the rotating frame) is negligible compared to the bulk $E_r \times B$ rotation---a common assumption in this context supported by both numerical and experimental evidence (see \cite{Schirmer_2006} and references therein)---then $v_\perp \approx E_r / B$ can serve as a proxy for the local radial electric field.
For the sake of generality, however, the results presented in this work are expressed in terms of $v_\perp$ rather than $E_r$, since the direct equivalence between the two is not guaranteed in general and still the subject of ongoing research\,\cite{Vermare_2024_IRW16, KraemerFlecken_2025}. Radial profiles of $v_\perp$ are obtained by stepping the probing microwave frequency, thereby varying the cutoff's radial location.\\
In principle, DBS provides access also to the turbulence wavenumber spectrum\,\cite{Hennequin_2006, Vermare_2011}, and---using multi-channel systems---to the instantaneous velocity shear\,\cite{Schirmer_2006}, turbulence radial correlation length\,\cite{Schirmer_2007}, or structure tilt angle\,\cite{Pinzon_2019a, Pinzon_2019b}. 
Although $v_\perp$ is the primary focus of this study, the current DBS setup also enables radial correlation studies on TCV, as recently demonstrated\,\cite{Panico_2024_EPS}.

\subsection{DBS launcher and operating range}\label{sec:launcher}
The DBS system is coupled to TCV's quasi-optical launcher antenna\,\cite{Cabrera_2018} shared with the short-pulse reflectometer (SPR)\,\cite{Cabrera_2019}. Hence, the DBS and SPR cannot operate simultaneously.
The antenna allows for toroidal and poloidal steering with \SI{0.2}{\degree} precision\,\cite{Cabrera_2018}. 
Furthermore, poloidal angle sweeps can be performed at a (safe) speed of $\lesssim \SI{30}{\degree \per \s}$.
The wave polarization can be switched flexibly between O- and X-mode. 
More technical details on the launcher can be found in~\cite{Cabrera_2018} and references therein.
\begin{figure}[htbp]
	\centering
		\includegraphics[width=1.0\linewidth]{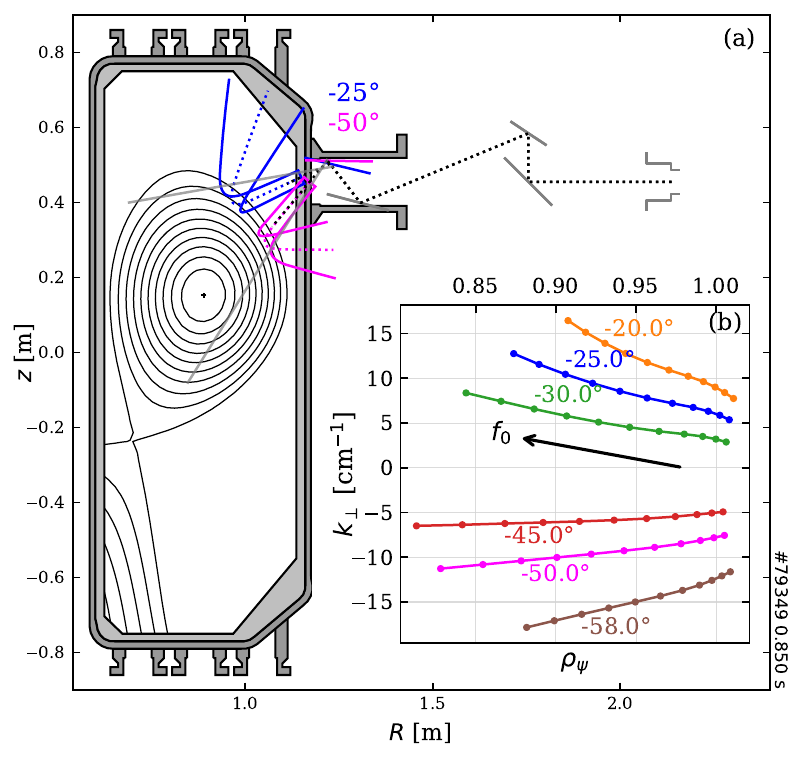}
	  \caption{(a) Illustration of the launcher geometry, showing two possible beam paths with opposite poloidal inclinations for an example TCV discharge. Gray solid lines indicate the accessible tilt angle range, $\theta \in [-58,-10]\SI{}{\degree}$. The inset (b) shows the reasonably accessible ($k_\perp,\rho_\psi$) space in this scenario. It is spanned by a set of 6 suitable poloidal angles and 11 equidistant probing frequencies ($f_0$) ranging from 48 to \SI{68}{GHz}.}\label{fig:launcher37}
\end{figure}
The beam is launched diagonally downward from the upper low-field side, as shown in \refig{launcher37}.
Using X-mode in typical L-mode conditions, the radial probing range lies between the separatrix and $\rho_\psi=0.7-0.9$,
 for low to high density scenarios at nominal $B$-field.
The probed wavenumber is usually contained within $k_\perp \sim [4,16]\,$cm$^{-1}$.
This translates roughly to $k_\perp \rho_s \sim [0.5, 2]$, typical of ion-scale turbulence ($\rho_s$ being the ion Larmor radius evaluated at the electron temperature). 
Typically, a sequence of 20 $\times \SI{5}{ms}$ frequency steps is programmed separately on both channels, to obtain a well-resolved (up to 40-points) profile of the mean $v_\perp$ flow every \SI{100}{ms}.

\subsection{Microwave scheme}\label{sec:microwave_scheme}
\begin{figure*}[htbp]
    \centering
    \begin{minipage}[b]{0.68\textwidth} 
        \centering
        \includegraphics[width=\textwidth]{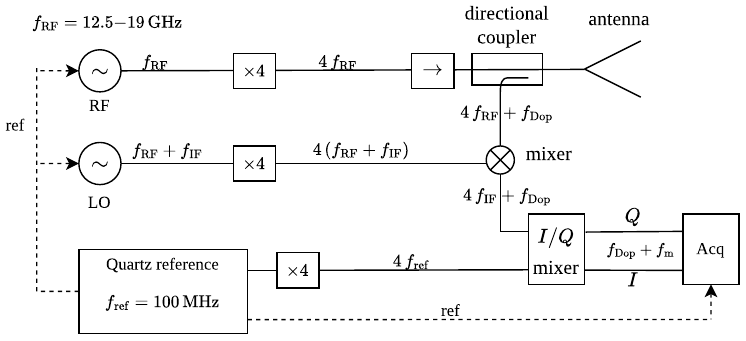} 
    \end{minipage}%
    \hfill
    \begin{minipage}[b]{0.3\textwidth} 
        \captionof{figure}{
            Simplified diagram of the microwave scheme used in the DBS hardware.
        }
        \label{fig:microwave_scheme}
    \end{minipage}
\end{figure*}

In principle, the microwave signal detection in the current system remains similar to the original implementation on Tore Supra\,\cite{Hennequin_2004}, with significant hardware upgrades introduced since.
A reduced diagram is shown in \refig{microwave_scheme}. 
For simplicity, only one of the two equivalent V-band channels is represented.
The scheme is based on heterodyne detection which uses a reference signal, or \enquote{local oscillator} (LO), in addition to the probing radio frequency (RF) signal. The RF and LO sources originate from two separate synthesizers that share a common phase reference provided by a quartz oscillator.
The LO signal is frequency-shifted by $f\low{IF}$ (typically $\SI{100}{MHz}$) with respect to the probing RF signal.
4-fold frequency multipliers raise the RF and LO signals to the V-band microwave range ($50$-$\SI{75}{GHz}$). The probing wave is then guided through a directional coupler---which separates the emitted from the received signals---to the antenna/launcher system. 

The subsequent demodulation consists of two steps. First, the received (backscattered) signal is mixed with the LO signal, retaining the low-frequency component oscillating at $4\,f\low{IF} + f\low{Dop}$, where $f\low{Dop}$ represents the Doppler shift. 
Second, it is mixed with the quartz reference signal at $4\,f\low{ref}=\SI{400}{MHz}$ through an in-phase and quadrature ($I$/$Q$) mixer.
The resulting $I$ and $Q$ signals are digitized and acquired with a 14 bits system at $10-\SI{100}{MHz}$ sampling rate.
In the conventional case of \textit{analog} $I$/$Q$ demodulation, the quartz and LO frequency offset are set equal, $f\low{ref} = f\low{IF}$. The acquired $I$ and $Q$ signals are thus left without a modulation frequency other than caused by the Doppler shift ($f_m = 0$).

An upgrade to this setup from \textit{analog} to \textit{digital} $I$/$Q$ demodulation has been performed recently, \added[id=A]{which is a distinguishing feature of the present DBS system.}
The idea is to retain a finite modulation frequency (here $f_m = \SI{6}{MHz}$) \replaced{at the output of the second ($I/Q$) mixer}{ in the $I/Q$ outputs} and to digitize the \added{$I$} signal with sufficient sampling rate to perform the final demodulation step numerically, instead of through the analog $I/Q$ mixer.
The practical advantages of this approach are twofold. 
First, analog $I/Q$ detection exhibits imperfections in amplitude balance and phase shift between the $I$ and $Q$ output signals, potentially distorting the resulting complex signal's power spectrum.
Second, digital demodulation bypasses the effect of a notch filter causing a broad spectral gap ($\Delta f \sim\SI{50}{kHz}$) around $f=0$. Especially small Doppler shifts are thus better resolved. 

\subsection{Data analysis procedure}\label{sec:data_analysis}

\subsubsection{Beamtracing.} 
The interpretation of DBS data requires knowledge of the radial coordinate $\rho$ and the wave number $k_i$ of the probing beam at its turning point. For a given measured Doppler shift $\omega\low{Dop}$, the turbulence velocity can then be localized and determined according to $v_\perp (\rho) = \omega\low{Dop} / k_\perp (\rho)$, where $k_\perp = -2 k_i$. To this end, we use the beamtracing code \texttt{beam3d}\,\cite{Honore_2006}, which was initially developed and remains in use for DBS at Tore Supra (now WEST).
The code propagates a set of interconnected rays representing the probing beam in 3D geometry, computing their trajectories within the \added[id=A]{Wentzel-Kramers-Brioullin (WKB, e.g.\,\cite[\S 2.8.1]{Hartfuss_2013})} approximation. The probing location is defined as the innermost radial point (in terms of $\rho_\psi$) along the beam trajectory, which usually coincides with the beam reflection point. 

\begin{figure*}[htbp]
    \centering
        \begin{subfigure}[b]{0.24\linewidth}
            \includegraphics[height=5.5cm]{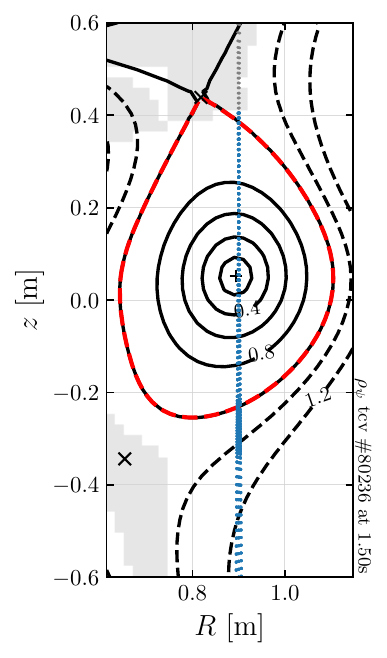}
            \caption{}\label{fig:}
        \end{subfigure}
        \begin{subfigure}[b]{0.39\linewidth}
            \includegraphics[height=5.5cm]{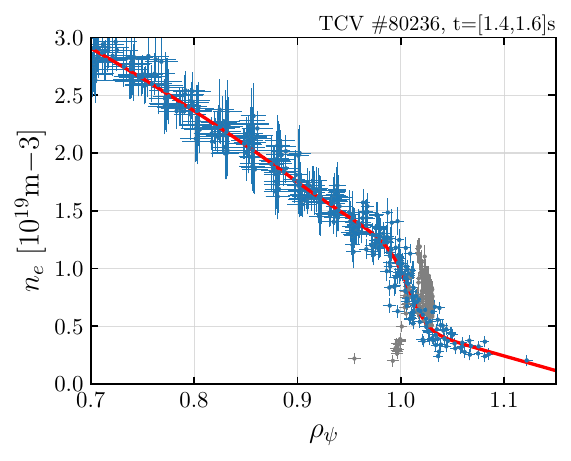}
            \caption{}\label{fig:}
        \end{subfigure}
        \begin{subfigure}[b]{0.34\linewidth}
            \includegraphics[height=5.5cm]{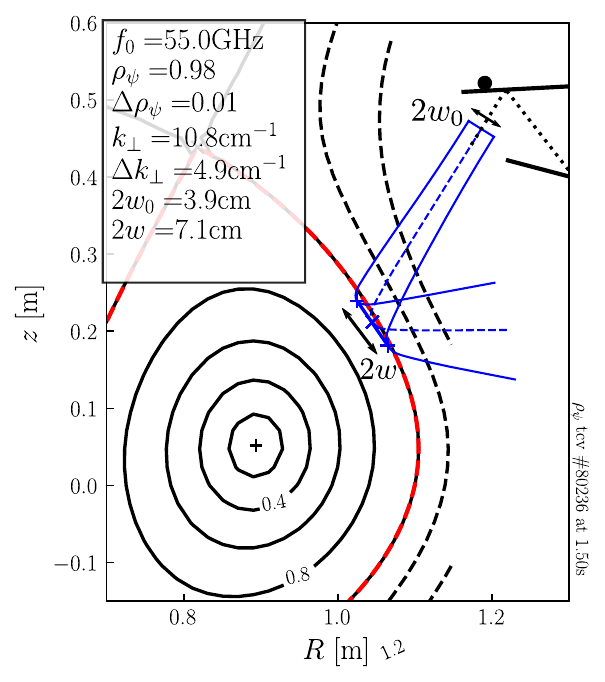}
            \caption{}\label{fig:}
        \end{subfigure}
        \caption{The beamtracing procedure is illustrated using an example TCV discharge: (a) Input magnetic equilibrium reconstruction and (b) density profile. The red curve represents a fit through the raw TS data, excluding spurious contributions (gray data points) stemming from the shaded regions in (a). The resulting beam trajectory is shown in (c) for a given probing frequency $f_0$.}\label{fig:TS_fit_example}
\end{figure*}

The interfacing with TCV input data is implemented as follows.
Beam properties (launch point, direction and waist radius) are retrieved from inherited routines shared with the SPR diagnostic\,\cite{Cabrera_2019}.
The magnetic equilibrium, \refig{TS_fit_example}\,(a), is obtained from the free boundary equilibrium reconstruction code LIUQE\,\cite{Moret_2015}.
The density profile, \refig{TS_fit_example}\,(b), is estimated from raw Thomson scattering (TS)\,\cite{Blanchard_2019} data.
Because of refraction, the beam trajectory and turning point location are sensitive to the density profile prior to the turning point. Accurately estimating the density profile across the separatrix and into the scrape-off layer (SOL) is therefore crucial to reach the desired level of precision in localizing the edge measurements.
This is achieved by fitting the edge TS data to a modified hyperbolic tangent (\textit{mtanh}) function (similar to $F\low{ped}$ in \,\cite{Stefanikova_2016}). A negative slope in the SOL ensures a smooth drop to zero to avoid any discontinuity in the refractive index.
TS measurements originating from regions close to the divertor legs or within the private flux region are excluded. These regions are shown in \refig{TS_fit_example}\,(a) as shaded areas. The corresponding data points, grayed out in \refig{TS_fit_example}\,(b), are systematically found to be outliers that would otherwise distort the pedestal fit.
\subsubsection{Doppler shift estimation.}
\begin{figure}[htbp]
     \centering
         \includegraphics[width=0.85\linewidth]{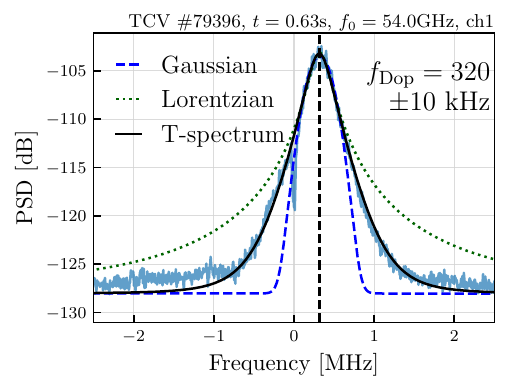}
     \caption{Example PSD and corresponding fits by a Gaussian, a Lorentzian and \added[id=A]{the test function (\enquote{T-spectrum}\,\cite{Hennequin_1999})}.}\label{fig:tcv_example_of_DBS_spectra}
 \end{figure}

In the present study, the $v_\perp$ profiles are inferred from DBS data acquired during stationary, MHD-quiescent, L-mode phases.
The mean Doppler shift can then be estimated directly from the power spectral density (PSD) through an appropriate peak finding method.
The PSD is estimated for each ($\approx\SI{5}{ms}$) frequency step using Welch's method ($n\low{FFT}=2048$ with 50\% overlap). A combination of Doppler peak estimators are used, from which the median value is retained. \replaced[id=A]{This includes Gaussian and Lorentzian curve fits, as well as the fit by a test function related to the statistical properties of turbulent motion\,\cite{Hennequin_1999}. The test function is denoted hereafter as the \enquote{T-spectrum}. It can describe the transition between a Gaussian and a Lorentzian function, reflecting either the ballistic or diffusive character of the underlying turbulent motion, respectively.}{This includes Gaussian and Lorentzian curve fits, as well as a T-spectrum\,\cite{Hennequin_1999} fit, which can be thought of as intermediate between a Gaussian and a Lorentzian function.} 
 An example is shown in \refig{tcv_example_of_DBS_spectra}\,(a).
By simultaneously minimizing the linear and logarithmic squared residuals, the T-spectrum fit is usually well constrained and captures both the peak and the wings of the distribution, which is generally not the case for Gaussian or Lorentzian fits.
In practice, spurious components associated e.g. to electronic noise have to be discarded to ensure an optimal fit. 
The uncertainty associated with the Doppler shift extraction
can be obtained by taking the lower and upper bounds across the different estimations. However, their scatter tends to be small for well-pronounced Doppler peaks, resulting in an underestimation of the uncertainty. Reasonable conservative uncertainties are therefore evaluated manually, based on the data quality and the width of the distribution. They are indicated as error bars for representative data points in the profiles shown.


\subsection{Verification of DBS against GPI results}\label{sec:DBS_vs_GPI}

TCV is equipped with a gas puff imaging (GPI) system\,\cite{Offeddu_2022} that can provide 2D data on the size and propagation of turbulent structures at different poloidal locations, from the SOL to just inside the separatrix. 
GPI diagnostics relies on the injection of a small amount of neutral gas---Helium in the discharges considered here---into the plasma edge.
The gas cloud interacts locally with the plasma, emitting visible light that is collected by a detector array or camera. Fluctuations in the measured brightness of a given spectral line are used to detect and track individual filamentary structures with high temporal resolution.
In particular, this enables the determination of their poloidal velocity.
Since \deleted[id=A]{the} DBS and GPI are fundamentally different measurement techniques, but should provide comparable information in terms of the background $E_r \times B$ flow carrying the turbulent structures, a cross-comparison of the two diagnostics is an excellent opportunity for an independent verification of the DBS results.

Here, the \textit{midplane} GPI system is used, which has a field of view located around the outboard midplane, spanned by a $12 \! \times \! 10$ detector array covering a $\SI{5}{cm} \!\times \! \SI{4}{cm}$ area in the radial and poloidal directions, respectively\,\cite{Offeddu_2022}. Conditional average sampling (CAS)\,\cite{Offeddu_2023}, combined with a tracking algorithm,
allows the average turbulence structure size and velocity to be estimated.
The average size of the detected structures varies depending on the plasma scenario and radial location, ranging from $\Delta_\theta \approx \SI{0.5}{cm}$ to
\SI{5}{cm} in the poloidal direction, or ${k_\perp} \sim \pi / \Delta_\theta  \approx 0.6$ to $\SI{6}{\per \centi \m}$.
The turbulence $k_\perp$ ranges accessed by the two diagnostics inside (and close to) the separatrix thus partly overlap, although the GPI tends to be more sensitive to the larger scales.
The helium gas is injected over $\SI{100}{ms}$ during a stationary discharge phase. The DBS data are analyzed for the same phase, but outside the puff injection period to minimize possible interferences. 

\begin{figure*}[htbp]
    \centering
    \begin{subfigure}[t]{0.7\linewidth}
        \includegraphics[width=\linewidth]{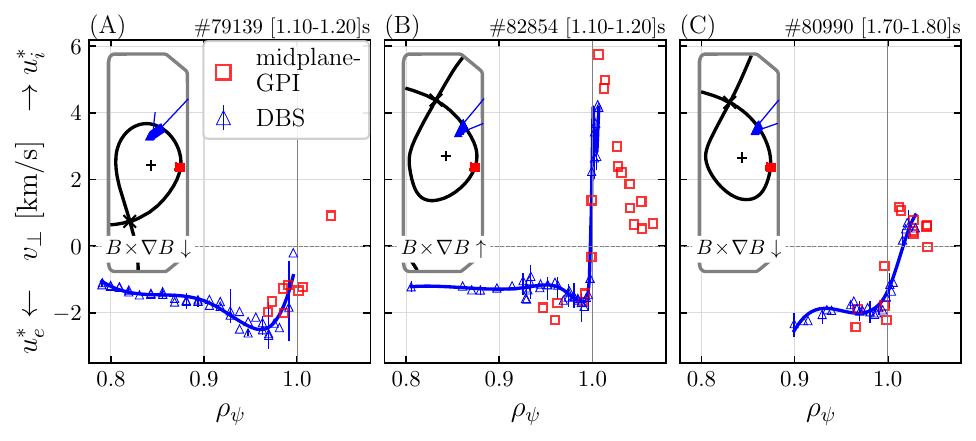}
        \caption{}\label{fig:DBS_GPI_comparison_a}
    \end{subfigure}
    \hfill
    \begin{subfigure}[b]{0.28\linewidth}

        \raggedright
            {\footnotesize
            \setlength{\tabcolsep}{3pt} 
                \begin{tabular}{c|ccc}
                \toprule
                  & Heating & $I_p$ [kA] & $\bar{n}_l^{19}$ [m$^{-3}$] \\ 
                \midrule
                (A) & Ohmic & $-210$ & 4.0 \\ 
                (B) & \makecell{ECRH\\(\SI{560}{kW})} & $180$ & 4.5 \\ 
                (C) & Ohmic & $-180$ & 6.1 \\ 
                \bottomrule
                \end{tabular}
            }
            \caption{}\label{fig:}
    \end{subfigure}
  \caption{(a) Comparison of the velocity profiles measured by the DBS and the midplane GPI diagnostics, respectively, in various plasma conditions and shapes (b) at fixed $B_0 = -\SI{1.44}{T}$. \added[id=A]{The DBS and GPI probing regions are highlighted in blue and red, respectively, in the plasma shape insets.} In (B), the GPI profiles have been shifted by $\Delta \rho_\psi = + 0.04$ to account for a clear radial mismatch with respect to the expected velocity reversal around the separatrix. }\label{fig:DBS_GPI_comparison}
\end{figure*}

The comparison is shown in \refig{DBS_GPI_comparison} for three different plasma conditions.
The distinction between perpendicular and poloidal projection of the velocity is considered small compared to the measurement uncertainties and therefore neglected here. In all cases, good agreement is found in the overlapping radial range, which lends confidence to both measurement techniques. 
Furthermore, it highlights the complementarity of the two diagnostics when it comes to measuring the full structure of the velocity across the separatrix, including the $E_r$ \enquote{hill} outside the separatrix.

\subsection{Comparison with the radial force balance}\label{sec:DBS_vs_CXRS}

To further validate the DBS measurements, we next proceed to confront them with the $E_r$ inferred from
the steady-state solution of the radial momentum conservation equation, or \textit{radial force balance},
\begin{equation}\label{eq:rad_force_balance}
E_r = V_{a, \varphi} B_\theta- V_{a, \theta} B_{\varphi} +\frac{\nabla_r p_a}{n_a e_a} \, .
\end{equation}
In this expression, the index $a$ denotes the considered species, $V_{a, \varphi}$ and $V_{a, \theta}$ are its fluid toroidal and poloidal velocities, respectively, $p_a = n_a T_a$ its pressure and $e_a$ its charge. 
In the following, we only consider the carbon (C$^{6+}$) impurity population and drop the species index.
The C$^{6+}$ profiles are measured via charge exchange recombination spectroscopy (CXRS\,\cite{Fonck_1984}).
In TCV, a low power, low torque diagnostic neutral beam (DNBI)\,\cite{Karpushov_2009}, provides the neutral donors for the CX reactions. The DNBI pulses typically cover a \SI{1}{s} period of the discharge with ON/OFF times set to 8/\SI{16}{ms} ($33\%$ duty cycle). The CXRS acquisition, synchronized with the DNBI blips, consists of four systems (SYS1-4)\,\cite{Marini_2017_phdthesis, Bagnato_2023}.
The toroidal velocity V$_{\varphi}$ can be obtained from the horizontal SYS1 and SYS2 systems at the low (LFS) and high field sides (HFS), respectively. Because of stronger attenuation of the neutral beam, the HFS system suffers from lower active signal compared to the LFS, and is therefore not considered here. $V_\varphi$ is thus obtained from SYS1.
The poloidal velocity V$_{\theta}$ can be measured at the low field side (LFS) with the vertical systems SYS3 and SYS4.
The latter was designed to provide highly resolved edge profiles\,\cite{Marini_2017_phdthesis}, but currently does not deliver reliable measurements of $V_\theta$. 
This leaves SYS3 as the sole system available for measuring $V_\theta$.
However, the data from SYS3 exhibits significant scatter, and the resulting profiles are sometimes questionable, especially towards the edge.
For a meaningful comparison with the DBS, we therefore resort to a neoclassical estimation \added[id=A]{as a proxy for} $V_\theta$ instead.
\added[id=A]{Though potentially inaccurate in certain plasma conditions or regions\,\cite{Solomon_2006, Tala_2007, Lebschy_2017}, past studies have found neoclassical estimates of the carbon poloidal flow to be in reasonable agreement with the measured one in L-mode TCV discharges\,\cite{Bortolon_2013}\cite[Fig.\,6.20]{Marini_2017_phdthesis}.} 
Here, the first order neoclassical poloidal flow of the carbon impurity is computed using the NEO code\,\cite{Belli_2008}, while the remaining terms on the right-hand side of \refeq{rad_force_balance} are estimated from spline fits to the CXRS data.
The diamagnetic term, $(\nabla p / Z n e)$, is inferred from SYS1 data.
All terms are divided by $B$, so that their sum, $v_{E_r \times B} = E_r / B$, is compared to the velocity $v_\perp$ measured by DBS.

\begin{figure*}[htbp]
	\centering
	\begin{subfigure}[b]{0.13\linewidth}
		\raggedleft
		\includegraphics[width=\linewidth]{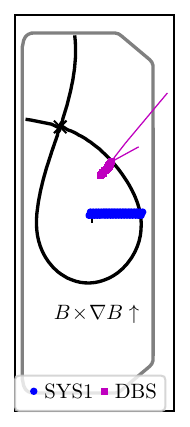}
		\caption{}\label{fig:DBS_vs_CXRS_eq}
	\end{subfigure}
	\begin{subfigure}[b]{0.44\linewidth}
		\centering
		\includegraphics[width=\linewidth]{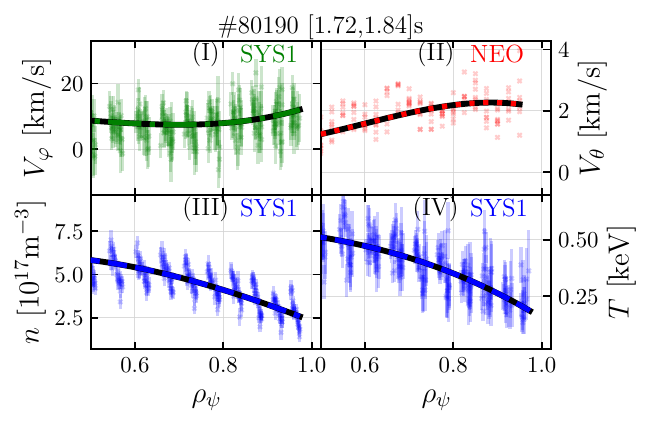}
		\caption{}\label{fig:DBS_vs_CXRS_splines}
	\end{subfigure}
	\begin{subfigure}[b]{0.41\linewidth}
		\raggedright
		\includegraphics[width=\linewidth]{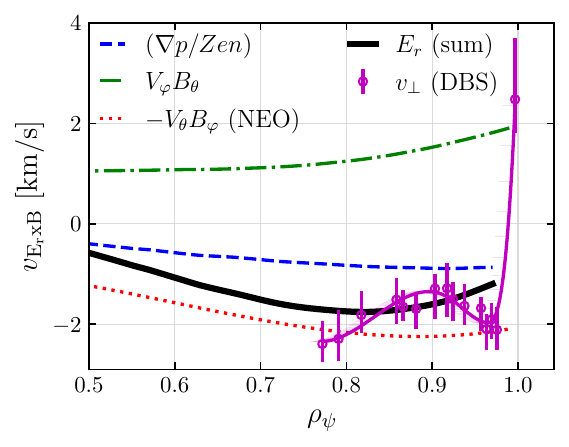}
		\caption{}\label{fig:DBS_vs_CXRS_splines_v}
	\end{subfigure}
	\caption{Comparison between DBS and radial force balance for C$^{6+}$, using CXRS measurements combined with a neoclassical prediction of $V_\theta$.
    (a) Equilibrium reconstruction of the considered scenario, showing the DBS and CXRS (SYS1) probing regions. 
    (b) Spline fits to the CXRS data (I,III,IV) and to the $V_\theta$ results from NEO (II). The scatter in $V_\theta$ is due to repeated NEO runs for each of the 12 CXRS acquisitions within the considered time window.
    (c) Radial force balance terms overplotted with the DBS $v_\perp$ profile.
	}\label{fig:DBS_vs_CXRS}
\end{figure*}

\refig{DBS_vs_CXRS} shows such a comparison for an example Ohmic L-mode discharge phase ($\#80190$ around \SI{1.8}{s}, USN, favorable $B\times \nabla B$, $B_0 = \SI{1.44}{T}$, $I_p=\SI{200}{kA}$, $\bar{n}_l=\SI{3e19}{\per \cubic \m}$).
In this particular example, a reasonable agreement is found regarding the general $E_r \times B$ velocity level derived from DBS and via \refeq{rad_force_balance}.
However, this quantitative agreement is not generally satisfactory in other cases considered. Moreover, the fine structure of $E_r$---in particular the well---is not captured by the radial force balance result.
On one hand, inaccuracies in the NEO-derived $V_\theta$ may stem from a deviation of the impurity flow from neoclassical predictions \added[id=A]{(as mentioned above)}, or uncertainties in the input data. On the other hand, the significant scatter in CXRS data, evident in \refig{DBS_vs_CXRS_splines} suggests large statistical uncertainties. Moreover, the spatial resolution of the CXRS systems may limit their ability to resolve small-scale corrugations in $E_r$.
We conclude that a quantitative analysis of the edge $E_r$ radial structure is difficult to complement by the CXRS measurements presently available. 
Nevertheless, using the analysis procedure detailed above, the CXRS data do not indicate any systematic or unreasonable disagreement with the DBS measurements.






\subsection{Mapping $v_\perp$ to the OMP}\label{sec:beam_incidence}

The steerable launcher allows various regions of the plasma to be probed, from the top to the outboard midplane (OMP), with positive and negative poloidal tilt angles relative to normal incidence.
Making use of this flexibility, we aim to investigate how the measured $v_\perp$ data is influenced by the beam poloidal tilt angle and/or by the poloidal probing location.
To begin with, we point out that the $E_r \times B$ flow not only varies poloidally with the magnetic field strength as $1/B$, but may also depend on the local flux expansion since the radial gradient of the electrostatic potential $\phi$ is weaker where flux surfaces lie further apart.
In vertically elongated plasmas (or in case of strong Shafranov shift), $E_r$ and the associated drift velocity should thus be weaker at the top than at the OMP, if the electrostatic potential $\phi$ is constant on a flux surface.

Following the example on AUG\,\cite{Hoefler_2021}, we proceed to map the $v_\perp$ data measured by DBS from different poloidal locations to the OMP, assuming that
it reflects the local $E_r \times B$ velocity, and that $\phi$ is a flux function.
Then, since $v_\perp = E_r /B = - \phi^\prime (\psi) \nabla \psi / B$, the local velocity measurement ($v_\perp$) should map to the OMP according to:
\begin{equation}\label{eq:v_perp_OMP}
    v_{\perp, \ttiny{OMP}} = v_\perp \frac{| \nabla \psi |_\ttiny{OMP}}{ | \nabla \psi | } \frac{B}{B_\ttiny{OMP}} \ ,
\end{equation}
where the flux expansion ratio $f_x / f_{x, \ttiny{OMP}} = | \nabla \psi |_\ttiny{OMP} / | \nabla \psi | $ can be significant when probing outside the OMP region in shaped TCV plasmas.
Here, $B$ varies moderately over the accessible probing range (typically $\lesssim{25}\%$), but since $B/B\ttiny{OMP}\ge 1$ (on a given flux surface), both the flux expansion and $B$-field variation tend to increase the magnitude of the velocity profiles when mapped to the OMP.



\begin{figure*}[htbp]
    \centering
    \begin{subfigure}[b]{0.41\linewidth}
        \includegraphics[height=5.5cm]{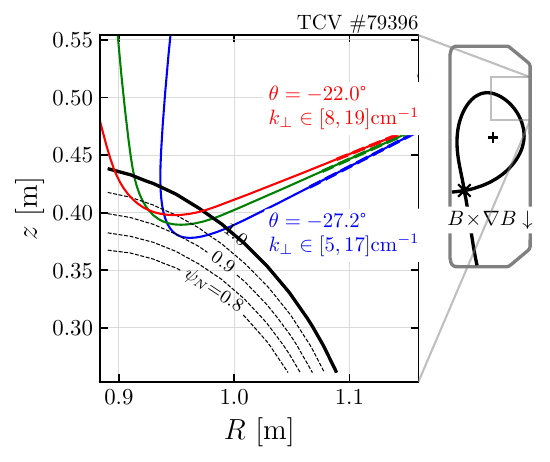}
        \caption{}\label{fig:}
    \end{subfigure}
    \begin{subfigure}[b]{0.58\linewidth}
        \includegraphics[height=5.5cm]{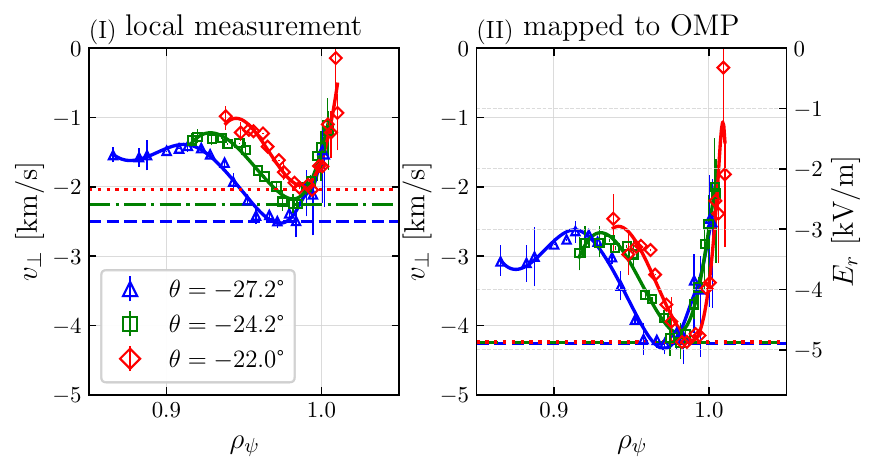}
        \caption{}\label{fig:}
    \end{subfigure}
  \caption{(a) Zoom onto the poloidal plane, showing the surfaces of constant normalized poloidal flux $\psi_N$, along with the computed ray trajectories at different tilt angles (for a given probing frequency). (b) Velocity profiles before (I) and after (II) mapping to the outboard midplane (OMP). The scale on the right-hand side represents the $E_r = v_\perp B$ values at the OMP, neglecting the radial variation of $B$ of a few \% over the probing range.}\label{fig:top_OMP_79396}
\end{figure*}

The mapping procedure is applied to $v_\perp$ profiles obtained at various beam poloidal angles within a fixed plasma scenario. First, we focus on the situation illustrated in \refig{top_OMP_79396}\,(a), where the beam is reflected towards the top (\textit{upward} probing), and the poloidal angle is swept from \SI{-28}{} to $\SI{-22}{\degree}$ over \SI{1.2}{s}.

The DBS profiles corresponding to three acquisition time windows along the sweep are displayed in \refig{top_OMP_79396}\,(b), confronting the velocities obtained (I) locally at the beam turning point with (II) the ones mapped to the OMP as per \refeq{v_perp_OMP}.
The mapping almost accounts for a factor 2 increase in the magnitude of the velocity (or $E_r$) well. 
Moreover, the initially differing well depths (dashed horizontal lines in \refig{top_OMP_79396}\,(b)) are brought together by the projection.
The shape of the profiles, on the other hand, including the radial location of the well, differs between the three cases, and the difference persists after the mapping.
The observed discrepancy (\replaced[id=A]{$\Delta \rho_\psi \sim 0.03$}{$\Delta \rho_\psi \sim 0.02$}) could stem from radial localization uncertainties associated with the input magnetic equilibrium or density profile used in the beamtracing. \deleted[id=A]{and/or systematic errors inherent to the beamtracing itself\,\cite{Conway_2019_IRW14}.}
\added[id=A]{Indeed, sensitivity studies revealed that slight differences in these input data could explain differences of the order of the observed radial shift.}
\added[id=A]{Additionally, systematic errors inherent to the beamtracing modelling are conceivable: At larger tilt angles relative to normal incidence (e.g., red trajectory in \refig{top_OMP_79396},a), both the spatial resolution and accuracy of the beamtracing tend to degrade\,\cite{Conway_2019_IRW14}.}
We conclude that in this particular case, the projection to the OMP according to \refeq{v_perp_OMP} yields
similar profiles---except for small radial shifts.

\begin{figure*}[htbp]
    \centering
    \begin{subfigure}[b]{0.41\linewidth}
        \includegraphics[height=5.5cm]{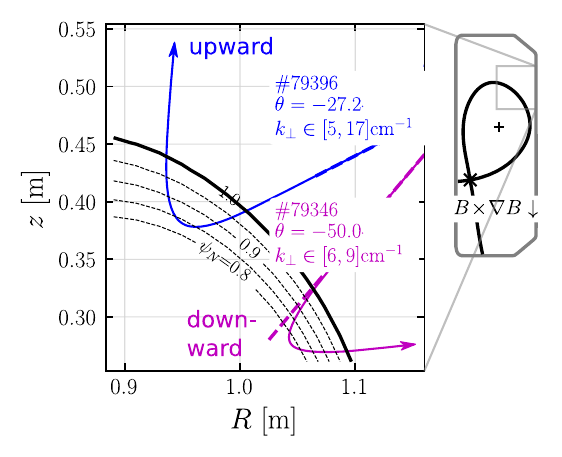}
        \caption{}\label{fig:}
    \end{subfigure}
    \begin{subfigure}[b]{0.58\linewidth}
        \includegraphics[height=5.5cm]{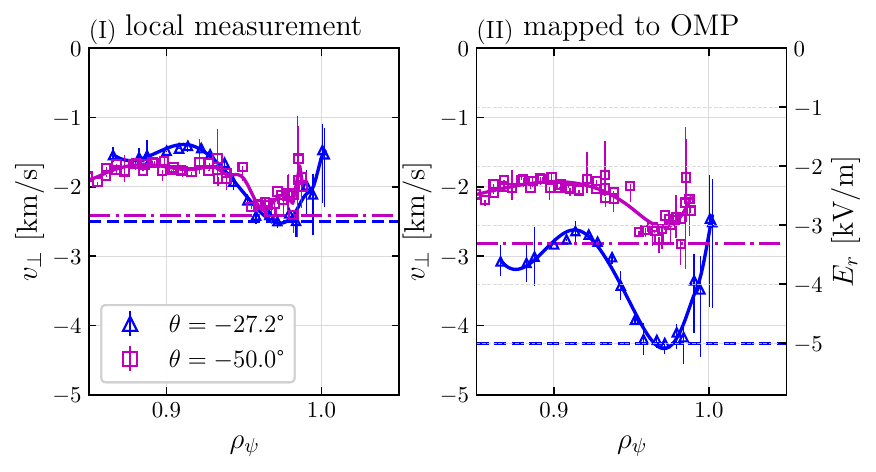}
        \caption{}\label{fig:}
    \end{subfigure}
  \caption{Same representation as in \refig{top_OMP_79396}, but with the comparison of up- and downward probing geometry. Here, the profiles agree well before (I), and disagree after mapping to the OMP (II).}\label{fig:top_OMP_79396_79346}
\end{figure*}

By contrast, the behavior is different when comparing the previous scattering geometry, that is, using \textit{upward} probing, with the \textit{downward} probing counterpart, see \refig{top_OMP_79396_79346}\,(a). A separate, well-matched discharge phase is used for this comparison.
Now, the \textit{local} $v_\perp$ profiles are similar, resulting in a large discrepancy once projected to the OMP, due to the combined effects---as mentioned previously---of higher flux expansion and lower $|B|$ in the top compared to the OMP region.
A qualitatively consistent behavior is observed in other scenarios (including in upper single-null discharges where the effect is enhanced due to the presence of the X-point), or in the same scenario but with reversed $B$-field and same helicity. 
Any hypothetical explanation for the discrepancy in terms of an effect that should reverse with the sign of $B$, therefore appears unlikely.
Moreover, the $k_\perp$ ranges probed in either beam direction partly overlap, as annotated in \refig{top_OMP_79396_79346}\,(a), ruling out any dominating influence of the selected structure sizes on the results.

A straightforward explanation for the results shown in \refig{top_OMP_79396_79346} is therefore not immediately apparent.
Possible causes that could be investigated in future work include: poloidal asymmetries in the electrostatic potential linked to neoclassical effects and/or ballooned turbulence, or diagnostic effects associated e.g. with eddy tilting.\footnote{In the presence of a non-zero mean tilt angle of the turbulent structures, the diagnostic response could be modified by 2D wave scattering effects depending on the probing direction relative to the structures tilt angle.}
Poloidal asymmetries in $v_\perp$ beyond what is implied by \refeq{v_perp_OMP} have been reported in other devices\,\cite{KraemerFlecken_2012,Vermare_2018, Estrada_2019}.
However, such poloidal asymmetry was not observed in a comparable study performed at AUG under conditions that appear closest to the present TCV scenarios, both in terms of DBS probing geometry as well as magnetic configuration\,\cite{Hoefler_2021}.

To conclude, when probing upward, the local $v_\perp$ measurements are scaled significantly if mapped to the OMP according to \refeq{v_perp_OMP}, leading to a possible inconsistency with the downward probing case. Hence, the DBS measurements obtained at different poloidal locations in TCV can presently not be considered equivalent. 
As a consequence, we renounce hereafter on the mapping procedure, presenting only \textit{local} $v_\perp$ measurements, obtained exclusively through \textit{downward} probing. The probing region is thus reasonably close to the OMP, where $f_x / f_{x, \ttiny{OMP}} \approx 1$ and where the validity of the DBS measurements has been confirmed by comparison with GPI, see \refig{DBS_GPI_comparison}.

\section{Characterization of $E_r$ sensitivities}\label{sec:results}

\subsection{Dependence on density}

We consider the evolution of $v_\perp$ across a density ramp performed in a pair of Ohmic discharges with comparable main parameters ($I_p \approx \SI{-170}{kA}$, $B_0 = \SI{-1.44}{T}$, $q_{95} \approx 3.8$--$4.0$, $\BxgradB$ pointing down), but different shapes: one limited, the other in diverted, upper single-null (USN) configuration.

\begin{figure}[htbp]
	\centering
	\includegraphics[width=1.0\linewidth]{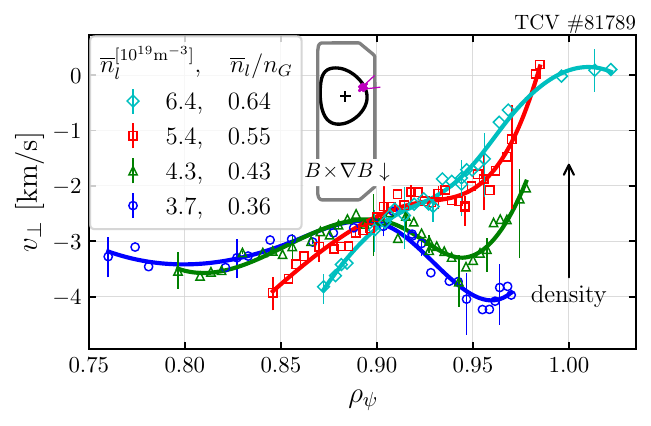}
	\caption{Density ramp in limited configuration: The well disappears with increasing density. The line-averaged density $\overline{n}_l$ is indicated in the legend along with the Greenwald fraction $\overline{n}_l/n_G$.}\label{fig:limited_dens_scan_81789_vperp}
\end{figure}
In the limited case, \refig{limited_dens_scan_81789_vperp}, 
the initially pronounced velocity well ($v_\perp \approx \SI{-4}{km/s}$) becomes shallower with increasing density, until it vanishes entirely at the highest density.
At the same time, an opposite trend is observed with respect to the inner (core) region of the profile: The velocity becomes more negative, which could be linked to a shift of the intrinsic toroidal velocity in the counter-current direction. Due to the lack of CXRS data for the limited discharge, this cannot be confirmed experimentally here, but it would be consistent with the reported trend of increasing counter-current rotation of the carbon impurity with density in limited Ohmic TCV discharges\,\cite{Bortolon_2006, Duval_2008}.\footnote{\added[id=A]{Note that the plasma current is well below the threshold ($I_p\approx\SI{290}{kA}$) above which intrinsic rotation reversals (from counter- to co-current) are expected to occur when exceeding a certain density in this configuration\,\cite{Duval_2008}.}}\deleted[id=A]{Note that the plasma current is well below the threshold ($I_p\approx\SI{290}{kA}$) above which intrinsic rotation reversals (from counter- to co-current) are expected to occur when exceeding a certain density in this configuration\,\cite{Duval_2008}.}

\begin{figure}[htbp]
	\centering
		\includegraphics[width=1.0\linewidth]{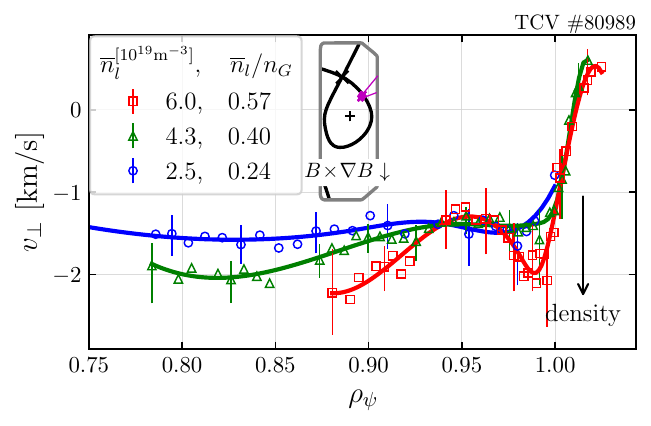}
	  \caption{Density ramp in diverted configuration ramp: The $v_\perp$ well becomes more pronounced with increasing density.}\label{fig:diverted_dens_scan_80989_vperp}
\end{figure}

By contrast, in the diverted case, \refig{diverted_dens_scan_80989_vperp}, 
the well only becomes clearly pronounced at the highest density\footnote{Although not entirely ruled out, a detachment under these discharge conditions seems unlikely, based on its absence during a comparable density ramp in LSN ($\#74400$) with better diagnostics coverage. This interpretation aligns with the observed non-vanishing $E_r$ peak in the SOL, indicative of attached divertor conditions\,\cite{Brida_2022}.},
as emphasized also by the appearance of a velocity \enquote{hill} to the inner side of the well, around $\rho_\psi=0.95$. This recurring \textit{inner} hill feature, which appears also in other TCV L-mode scenarios, is particularly visible here at high density. 
It is reminiscent of a similar observation made on WEST for a comparable magnetic configuration (USN, unfavorable $\BxgradB$)\,\cite{Peret_2022_phdthesis}.
The density is thus found to affect the $E_r$ well structure in a qualitatively different way depending on the
magnetic boundary condition and/or shape. 
In the core, however, the trend appears consistent between the limited and diverted case. 
A deepening of the $E_r$ well with density has been documented in JET\,\cite{Silva_2024_submitted} and in AUG\,\cite{Plank_2023}. General conclusions on the density (or collisionnality) dependence of $E_r$ in TCV, however, require systematic investigations over a larger set of plasma conditions, including favorable $\BxgradB$ drift, which is part of ongoing work.

\subsection{Dependence on auxiliary heating}

\begin{figure}[htbp]
	\centering
	\includegraphics[width=1.0\linewidth]{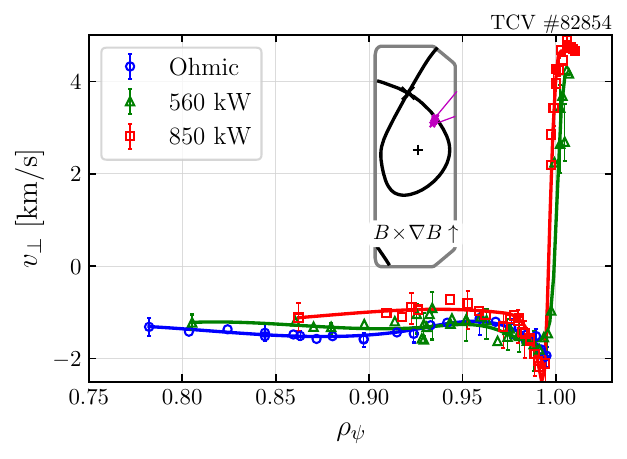}
	\caption{ECRH power scan: The flow profile shows little variation with increasing X2 heating.}\label{fig:ecrh_82854_steps_vperp}
\end{figure}

\begin{figure}[htbp]
	\centering
	\includegraphics[width=1.0\linewidth]{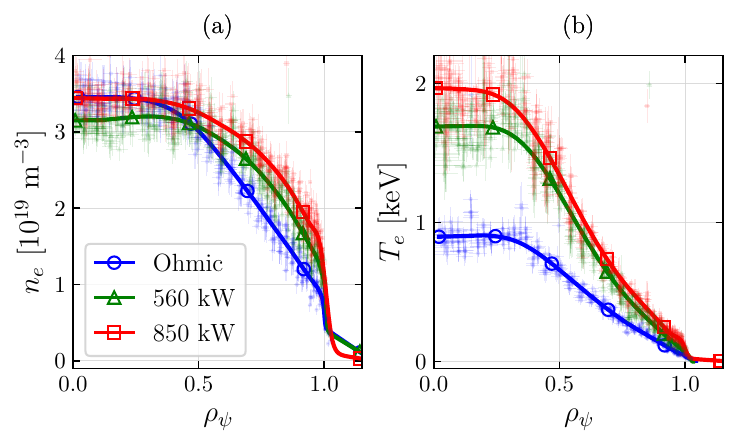}
	\caption{Electron (a) density and (b) temperature profiles across the ECRH power scan. The curves represent fits to Thomson scattering data, while the different marker symbols are used solely for distinction.}\label{fig:ecrh_82854_steps_kinetic_profs}
\end{figure}

We proceed to investigate the impact of auxiliary heating on the $E_r$ profile, starting with electron cyclotron resonance heating (ECRH),
followed by neutral beam injection (NBI).
In the former case, stepped power increments are applied to an USN, favorable $\BxgradB$ discharge with $I_p \approx \SI{170}{kA}$, $B_0 = \SI{1.44}{T}$, $q_{95} \approx 4.2$ and \replaced{$\overline{n}_l =2.2$--$\SI{2.9e19}{\per \m \cubed}$}{$\overline{n}_l =4$--$\SI{5e19}{\per \m \cubed}$}.
Second harmonic (X2) ECRH is used, with central deposition and complete absorption inside $\rho_\psi \lesssim 0.4$, according to linear raytracing performed with the TORAY-GA code\,\cite{Kritz_1982}.
The initial Ohmic phase and two subsequent steps of increasing ECRH power are examined over stationary time windows.
The evolution of the $v_\perp$ profile is shown in \refig{ecrh_82854_steps_vperp}, along with the kinetic profiles in \refig{ecrh_82854_steps_kinetic_profs}.

The direct electron heating is evident from the increase in electron temperature ($T_e$) shown in \refig{ecrh_82854_steps_kinetic_profs}\,(b). Also the edge density increases with the ECRH power, \refig{ecrh_82854_steps_kinetic_profs}\,(a). This is accompanied by the formation of a pronounced L-mode pedestal and an \deleted{mild} increase of the line-averaged density by \replaced{$<\SI{30}{\%}$}{$<\SI{20}{\%}$} across the scan.
By contrast, the $v_\perp$ profiles, \refig{ecrh_82854_steps_vperp}, show little variation with $P\low{ECRH}$ inside the LCFS, neither in the well region nor deeper in the core. 
The carbon impurity profiles measured by CXRS (not shown) do not indicate any significant change in the \added[id=A]{core ion temperature ($T_{C^{6+}} \approx 0.5 \pm \SI{0.1}{keV}$ around $\rho_\psi \approx 0.25$)} between the Ohmic and \SI{850}{kW} ECRH phase. Only a mild increase by $30\%$ or $\Delta T_{C^{6+}} \approx\SI{50}{eV}$ is found in the outer core region ($\rho_\psi \in [0.7,0.9]$).
Finally, note that \textit{outside} the LCFS, the velocity \enquote{hill} \replaced[id=B]{in the presence of ECRH is notably higher ($v_\perp > \SI{4}{km/s}$) than for instance in the green and red profiles shown in \refig{diverted_dens_scan_80989_vperp} ($v_\perp \lesssim \SI{1}{km/s}$), which correspond to a comparable equilibrium at higher density and without ECRH (hence cooler edge)}{reaches high values in the presence of ECRH ($v_\perp > \SI{4}{km/s}$), compared to pure Ohmic heating ($v_\perp \lesssim \SI{1}{km/s}$, see \refig{diverted_dens_scan_80989_vperp})}.\footnote{\added[id=B]{The reduction in SOL $E_r$ peak with reduced electron heating and/or increasing density is also evident from comparing the GPI data in \refig{DBS_GPI_comparison_a} (B) and (C).}}
This is consistent with an increased $T_e$---and associated gradient---across the LCFS in the ECRH case, leading to higher $E_r$ values in the near SOL according to the expected $E_r^\text{SOL} \sim - \nabla_r T_{e, \ttiny{OMP}} / e$ scaling\,\cite{Stangeby_2000, Brida_2022}.

\begin{figure*}[htbp]
	\centering
	\includegraphics[width=.8\linewidth]{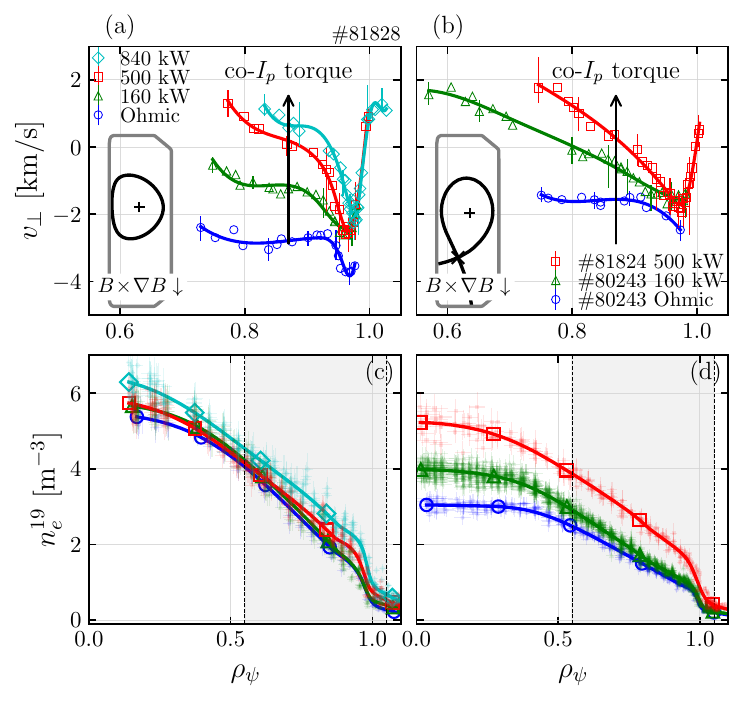}
	\caption{Co-current injected NBI power steps in (a) inner wall limited and (b) diverted configuration, for otherwise comparable plasma conditions. (c) and (d) show the corresponding density profile fits, respectively. Note the different radial ranges displayed in the upper and lower graphs.}\label{fig:NBI_steps_lim_diverted}
\end{figure*}

Next, the impact of tangential neutral beam injection (NBI) is investigated in a set of discharges covering both limited and diverted configurations. The first of two available NBI systems is used (NBI-1, \SI{28}{keV} deuterium, $P\low{NBI} \le \SI{1.3}{MW}$\,\cite{Karpushov_2023}) with beam injection in the co-current direction for the discharges considered.
In addition to the preferential ion heating, the toroidal torque associated with the NBI is expected to drive changes in $E_r$, since according to the ion radial force balance, \refeq{rad_force_balance}, a co-$I_p$ (counter-$I_p$) toroidal momentum increment tends to increase (reduce) $E_r$.
 \refig{NBI_steps_lim_diverted} shows the evolution with stepped power increments, for (a) limited and (b) diverted L-mode discharges, respectively.
Apart from the different magnetic configuration and shape, and a higher initial density ($\overline{n}_l=\SI{3.3e19}{\per \m \cubed}$) in the limited compared to the diverted case ($\SI{2.5e19}{\per \m \cubed}$), the macroscopic plasma parameters are otherwise similar ($I_p = \SI{-170}{kA}$, $B_0 = \SI{-1.44}{T}$, $q_{95} \approx 4$).\\
In both cases, the core $v_\perp$ profile shifts upwards with increasing NBI power, and reverses 
the sign increasingly closer to the edge.
This is qualitatively consistent with the expected $E_r$ increment in response to the co-$I_p$ torque, as discussed above.
The first power step leads to a notable reduction in the magnitude of the $E_r$ well compared to the Ohmic phase. 
\deleted[id=A]{This could be due to the significant toroidal momentum change in the core caused by the onset of NBI heating, which drives the initially weak intrinsic toroidal rotation into a strong co-$I_p$ rotation, an effect likely to extend partially to the edge region.}
The subsequent power increments do not lead to any further significant evolution of the well. The velocity \textit{shear}, on the other hand, keeps increasing as the inner branch of the profile shifts upwards.\\
The overall density and that of the pedestal increase with the NBI power---especially in the diverted case, \refig{NBI_steps_lim_diverted}\,(d). This could be due to a neutral beam fueling effect and/or increased particle confinement. Thus, there is inevitably some level of mixing between the effect of increased density and NBI power on the rotation and $E_r$.
Systematic DBS measurements in dedicated \textit{counter}-$I_p$ NBI scans have not been performed yet. While the expected opposite trend ($v_\perp$ becomes more \textit{negative} towards the core) has been observed for increasing counter-$I_p$ torque, this is yet to be confirmed and documented over a broader range of discharge conditions and NBI powers.\\

\subsection{Magnetic field helicity}

\begin{figure*}[htbp]
	\centering
    \begin{subfigure}[b]{0.2\linewidth}
        \raggedleft
        \includegraphics[height=7.cm]{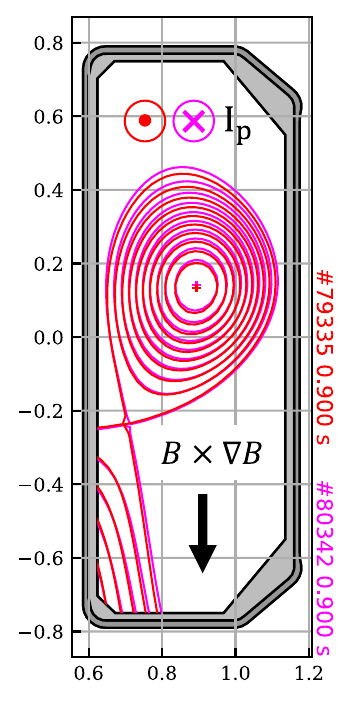}
        \caption{}\label{fig:}
    \end{subfigure}
    \begin{subfigure}[b]{0.58\linewidth}
        \centering
        \includegraphics[height=7.5cm]{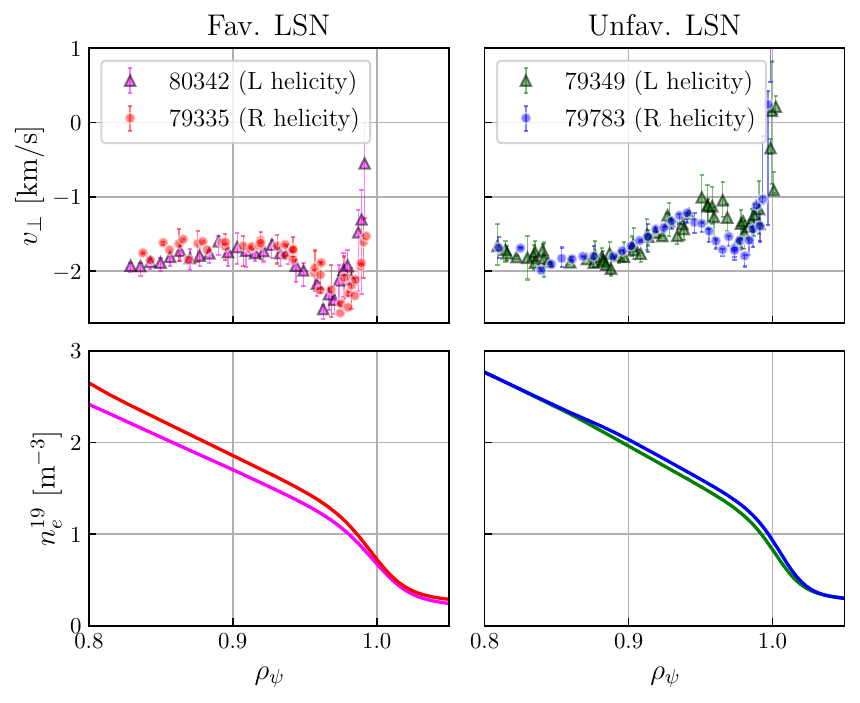}
        \caption{}\label{fig:}
    \end{subfigure}
    \begin{subfigure}[b]{0.2\linewidth}
        \raggedright
        \includegraphics[height=7.cm]{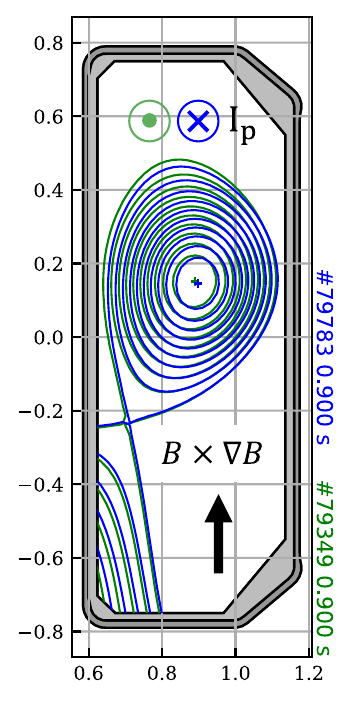}
        \caption{}\label{fig:}
    \end{subfigure}
    \caption{$B$-field direction and helicity scan at $I_p = \pm \SI{210}{kA}$ and $B_0 = \pm \SI{1.44}{T}$: (a) and (c) Equilibrium reconstructions for Fav. and Unfav. $\BxgradB$ drift, respectively. (b) Measured velocity profiles for the four discharges, showing good agreement between \enquote{right-handed} (R) and \enquote{left-handed} (L) helicity for a given $\BxgradB$ configuration.
    }\label{fig:helicity_scan_tcv}
\end{figure*}

The $E_r$ well is empirically known to be sensitive to the magnetic topology, in particular to the ion $\BxgradB$ drift with respect to the active X-point\,\cite{Schirmer_2006, Vermare_2021, Plank_2023}.
A systematic analysis of the favorable/unfavorable $\BxgradB$ dependence of $E_r$ in TCV is ongoing and will be presented elsewhere.
Here, we focus on the effect of flipping the magnetic field line helicity, by comparing the four possible sign permutations of $I_p$ and $B_0$ in a fixed LSN scenario.
By symmetry, the latter might not be expected to play a significant role. However, possible diagnostic biases, or symmetry-breaking effects associated with plasma-wall interactions, could still influence the measurements, motivating a verification.
\refig{helicity_scan_tcv} shows the results for a quartet of matched discharges with $I_p= \pm \SI{210}{kA}$ and $B_0 = \pm \SI{1.44}{T}$.
For a given drift configuration, no significant difference in $v_\perp$ is observed when flipping the helicity. The slight differences outside the errorbars in the reversed (unfavorable) $\BxgradB$ case could be due to an imperfect match in plasma \added[id=A]{conditions}\footnote{\added[id=A]{The shapes do not exactly overlap between L and R helicity, and the core $T_e$ in 79783 is lower by 10-15\% compared to the remaining discharges. The core density, on the other hand, is well-matched between 79783 and 79349, while it differs by 10-15\% in the favorable pair.}}, \added[id=A]{or an actual effect related to the helicity---albeit small}. The relative insensitivity of the DBS profiles to the helicity sign were also confirmed in a similar comparison at $I_p=\SI{150}{kA}$ and otherwise same conditions.
While it is not guaranteed to be generally the case, e.g. in density, temperature or magnetic equilibrium
we conclude that the helicity sign can be freely chosen for this particular scenario in future studies. A practical implication is that forward/reversed $B$ (or favorable/unfavorable $\BxgradB$) configurations may thus be meaningfully \replaced[id=B]{confronted even}{ compared also} when using NBI heating, keeping the direction of $I_p$ fixed to use the \textit{same} neutral beam source---an important consideration given that the two neutral beams in TCV are not strictly comparable.

\section{Summary and conclusions}\label{sec:conclusion}

We have presented the commissioning of a dual channel V-band DBS system in TCV 
that has been employed to survey the edge $E_r$ radial profile and its sensitivities in this tokamak.
The study focussed on L-mode discharges, where X-mode polarization is used to probe the edge region, typically ranging from the LCFS to $\rho_\psi \approx 0.7$-$0.9$.
In standard operation, a 40-point profile of $v_\perp \approx E_r \times B$ is acquired every \SI{100}{ms}.
Digital demodulation---a distinguishing feature of the present DBS system---allows for high accuracy measurements even in the case of small Doppler shifts.
Cross-comparison between the DBS and GPI diagnostics shows good agreement, and the measured level of $E_r$ is found compatible with the one inferred from the impurity radial force balance.
Yet, attempts to map local $v_\perp$ measurements from different poloidal beam orientations onto the outboard midplane remain partly inconclusive. Consequently, the mapping was not applied, and the probing beam was maintained in a downward tilt configuration throughout the subsequent study.
The $E_r$ radial profile was examined across scans in density, auxiliary heating, and magnetic field helicity. In diverted configuration, the depth of the velocity well just inside the LCFS typically reaches $v_\perp \approx -\SI{2}{km \per \s}$ (or $E_r \approx -\SI{2.5}{kV \per \m}$) with moderate variation across the scanned parameters.
In limited configuration, the velocity well is relatively deep ($v_\perp \approx -\SI{4}{km/s}$) at low density, and becomes significantly shallower as density increases.
NBI consistently affects the outer core $E_r$ profile in both limited and diverted scenarios, aligning qualitatively with the expected $E_r$ response to the toroidal torque.
In contrast, ECRH shows a comparably small impact. Finally, no significant sensitivity of the $E_r$ well to the sign of the magnetic field helicity is observed.
The influence of magnetic shaping and of the $B \times \nabla B$ drift on $E_r$ remains to be investigated in TCV.\\

\appendix 

\section{Coordinates and sign conventions}\label{append:conventions}

This paper adopts the standard toroidal coordinates convention for TCV, COCOS17\,\cite{Sauter_2013}, so that a positive (negative) toroidal magnetic field or plasma current is directed counterclockwise (clockwise) when viewed from above.
By contrast, the sign convention used for the velocity profiles inferred from DBS, is chosen such that $v_\perp>0$ ($v_\perp <0$) corresponds to a \textit{perpendicular} rotation (that is, binormal to $B$ and $\nabla \psi$) in the ion (electron) diamagnetic direction. Thus, $v_\perp$ and the associated radial electric field $E_r$ (assuming equivalence between $v_\perp$ and $v_{E_r \times B}$) have the same sign regardless of the $B$-field orientation.
Radial coordinates are expressed in terms of the normalized flux surface label $\rho_\psi = \sqrt{(\psi - \psi\low{0}) / (\psi\low{LCFS} - \psi\low{0})}$, where $\psi\low{0}$ and $\psi\low{LCFS}$ are the poloidal magnetic fluxes at the magnetic axis and at the last closed flux surface (LCFS), respectively.

\ack

The authors would like to thank Y. Camenen and A. Merle for their assistance with the NEO simulations, and X. Garbet and O. Février for helpful insights.\\
This work has been carried out within the framework of the EUROfusion Consortium,
funded by the European Union via the Euratom Research and Training Programme
(Grant Agreement No 101052200---EUROfusion). Views and opinions expressed
are however those of the author(s) only and do not necessarily reflect those of the
European Union or the European Commission. Neither the European Union nor the
European Commission can be held responsible for them.\\
This work has benefited from a grant managed by the Agence Nationale
de la Recherche (ANR), as part of the program \enquote{Investissements
d'Avenir} under the reference (ANR-18-EURE-0014).\\
This work was supported in part by the Swiss National Science Foundation.

\printbibliography[nottype=urlyear]

\end{document}